\documentclass[review]{elsarticle}

\usepackage{amsmath}
\usepackage{hyperref}

\journal{Chaos, Solitons \& Fractals}

\begin{document}

\begin{frontmatter}

\title{Creating New Chaotic Signals with Reservoir Computers}

\author[mymainaddress]{Thomas L. Carroll \corref{mycorrespondingauthor}}
\ead{thomas.carroll@nrl.navy.mil}

\address[mymainaddress]{Code 6392, US Naval Research Lab, Washington DC 20375 USA.}

\begin{abstract}
While there have been many publications on potential applications of chaos to fields such as communications, radar, sonar, random signal generation, channel equalization and others, designing continuous chaotic systems is still an unsolved problem. There are a number of well known chaotic systems used for applications, but if any application is to become widely used, some way of generating many different chaotic signals is necessary. This work shows that one may use a reservoir computer to create a set of chaotic signals that are correlated but easily distinguishable from one chaotic signal with desirable properties. The ability to distinguish the new signals is demonstrated with a simple communications example.
\end{abstract}


\end{frontmatter}


\section{Introduction}
While much has been published on the topic of using chaos for communications\cite{cuomo1993a,parlitz1992, hayes1994,  yang1997, abel2002, argyris2005, shanmugam2006, blakely2013, kaddoum2016a,bai2020} or radar \cite{ wu2001,  Hara2002, Machowski2002, Lin2004,carroll2007, gambi2008, rachford2010, blakely2011,pappu2020}  , one problem that has not been much addressed is the design and implementation of chaotic systems for these applications. There have been a number of chaotic systems proposed for these uses, such as the Lorenz \cite{lorenz1963} or R{\" o}ssler \cite{rossler1976} systems, the Chua system \cite{matsumoto1985}, the 19 Sprott systems \cite{sprott1994} and others, but more alternatives are needed for actual applications. There are no rules for designing chaotic signals to have desired properties; Sprott does give a list of chaotic systems, but these were found by a systematic search. Adding to the design complexity, in some cases one may want to use the self synchronizing property of chaos, but the chaotic system must be designed specifically to allow this possibility. One may also create different signals from a known chaotic system by changing a parameter, but there may be limits on how far the parameter may be changed without encountering a bifurcation.

In this work we propose a method to create a large number of chaotic signals from a particular chaotic system that has desirable properties. We use a chaotic signal from a desirable system such as a Lorenz system to drive a reservoir computer. A reservoir computer is a high dimensional dynamical system that may be created by connecting a number of nonlinear nodes in a recursive network \cite{jaeger2001,natschlaeger2002} . Usually the output signals from this network are combined to fit a training signal; for our purposes, we instead make random combinations of signals from the reservoir to create a new set of signals. These signals are nonlinear functions of the original driving signal; while they are still correlated with the driving signal, they can still be distinguished from the driving signal and each other by training a second  reservoir computer on the new signals. We show that these signals may be used to communicate using chaos shift keying (CSK), in which different communications symbols are represented by signals from different chaotic systems.

One feature of reservoir computers is that because the training only takes place on the output, they may be constructed from analog systems. Reservoir computers that are all or part analog include photonic systems \cite{appeltant2011,larger2012, van_der_sande2017, hart2019,chembo2019,argyris2019}, analog electronic circuits \cite{schurmann2004}, mechanical systems \cite{dion2018} and field programmable gate arrays \cite{canaday2018}. Many other examples are included in the review paper \cite{tanaka2019}. Building reproducible analog chaotic circuits that operate at high frequencies or high powers is difficult, so one could envision driving an analog reservoir computer with a digital signal to produce a number of analog chaotic signals.

In this work, a reservoir computer will be used to create a new set of chaotic signals from an input signal. A second reservoir computer will be trained on each of these new signals and the training coefficients for each new signal will be stored. To transmit information, for each data interval, one of these new signals will be transmitted. The job of the receiver is to determine which of these signals was sent for each data interval. Two types of receiver are studied; one where the receiver is synchronized to the original chaotic signal in the transmitter and one in which it is not synchronized.

\section{Reservoir Computers}
The reservoir computer we use in this work, often known as the leaky hyperbolic tangent reservoir computer \cite{jaeger2001}, is common in the literature. It is described by
\begin{equation}
\label{tanhnode}
{\bf {R}}\left({n+1}\right)=\left(1-\alpha\right)\mathbf{R}\left(n\right)+\alpha\tanh\left({{\bf {AR}}+\mathbf{W}^{in}s\left(n\right)}+1\right)
\end{equation}
where ${\bf R} $ is a vector of reservoir variables, ${\bf A}$ is the adjacency matrix that describes how the different nodes are connected, $s$ is the input signal and ${\bf W}^{in}$ is the vector of input coefficients. The individual components of ${\bf R}(n)$ are $r_i(n)$, where $i$ is the index of a particular node. The reservoir computer has $M$ nodes, so the dimensions of ${\bf R}$ and ${\bf W}^{in}$ are $M \times 1$ while ${\bf A}$ is $M \times M$.

 In the training stage, the reservoir computer is driven with the input signal $s(n)$ to produce the reservoir computer output signals $r_i(n)$. In all the examples in this paper, the input signal is normalized to have a mean of zero and a standard deviation of 1. The reservoir output matrix $\Omega_1$ is constructed from the reservoir signals as
\begin{equation}
\label{newmat}
\Omega_1=\left[{\begin{array}{cccccc}
{{r_{1}}\left(1\right)} & \cdots & {{r_{M}}\left(1\right)}\\
{{r_{1}}\left(2\right)} & {} & {{r_{M}}\left(2\right)}\\
\vdots & {} & \vdots\\
{{r_{1}}\left(N\right)} & \cdots & {{r_{M}}\left(N\right)}
\end{array}}\right]
\end{equation}

\subsection{Creating New Signals}
\label{resnew}
In normal use the a linear combination of the columns of the matrix $\Omega_1$ would be used to fit a training signal. Instead, to create $N_s$ new signals, we create a $M \times N_s $ random matrix of coefficients $\Psi$. The elements of $\Psi$ are drawn from a uniform random distribution between -1 and 1. To insure that the columns of $\Psi$ are not too similar to each other, then are then made orthonormal to each other by a Gram-Schmidt or other method to yield $\Psi_O$. We produce $N_s$ new signals as
\begin{equation}
\label{randsig}
\Theta=\Omega_{1}\Psi_O .
\end{equation}

The new signals are $\Theta_j(n), n=1 \ldots N, j=1 \ldots N_s$, where $N$ is the number of points in the reservoir time series.

\subsection{Distinguishing the New Signals}
A second reservoir computer may be used to distinguish the different chaotic signals. The reservoir computer is driven with one of the signals from the matrix of signals $\Theta$:
\begin{equation}
\label{resdid}
{\bf {R_{j}}}\left({n+1}\right)=\left(1-\alpha\right)\mathbf{R_{j}}\left(n\right)+\alpha\tanh\left({{\bf {AR_{j}}}+\mathbf{W}^{in}\Theta_j\left(n\right)}+1\right)
\end{equation}
Before driving, each signal in $\Theta$ is normalized by subtracting the mean and dividing by the standard deviation.

The output signals from the reservoir computer driven with each of the new signals are each arranged in a matrix
\begin{equation}
\label{fitmat}
\Omega_{j}=\left[{\begin{array}{cccccc}
{{r_{1j}}\left(1\right)} & \cdots & {{r_{Mj}}\left(1\right)} & 1\\
{{r_{1j}}\left(2\right)} & {} & {{r_{Mj}}\left(2\right)} & 1\\
\vdots & {} & \vdots & \vdots\\
{{r_{1j}}\left(N\right)} & \cdots & {{r_{Mj}}\left(N\right)} & 1
\end{array}}\right]
\end{equation}
where the first index of $r$ indicates the node number. The last column of $\Omega_j$ is set to 1.0 to fit any constant offset.

The reservoir computer is trained on a particular chaotic signal, the reservoir is trained by predicting that signal one time step into the future. The training signal is $g_j(n)=\Theta_j(n+1)$. For each of the $N_s$ new signals, the matrix $\Omega_j$ is used to fit the training signal as
\begin{equation}
\label{trainfit}
g_j\approx h_j=\Omega_{j}{\bf W}^{out}_{j}
\end{equation}
where the fit is done using ridge regression to prevent overfitting. The fit coefficients are in the vector ${\bf W}^{out}$. The training error $\Delta_j^{RC}$ is the standard deviation of $g_j-h_j$, normalized by the standard deviation of $g$.

For signal identification, the reservoir computer of eq. (\ref{tanhnode}) is driven with a signal ${\tilde  s}$ from the same dynamical system with different initial conditions. The output signals ${\tilde {\bf R}}$ are arranged in a matrix ${\tilde \Omega}_1$ and a set of new signals is created as ${\tilde \Theta}={\tilde \Omega}_1 \Psi_O$. The testing signals are ${\tilde g}_j(n)={\tilde \Theta}_j(n+1)$, which may be approximated as ${\tilde h}_j={\tilde \Omega}_{j}{\bf W}^{out}_{j}$. The testing error $\Delta_j^{tx}$ is the standard deviation of ${\tilde g}_j - {\tilde h}_j$.

\section{Communications: Synchronous and non-Synchronous}
The different signals $\Theta_j, j=1 \ldots N_s$ may be used as communications symbols, an encoding commonly known as chaos shift keying (CSK).  Typically CSK would proceed by sending signals from different chaotic systems (also known as attractor shift keying) or by sending signals from one chaotic system with different parameters. The version of CSK described here is equivalent to sending different components from a chaotic system. 

The communications system may be divided into signal encoding and signal decoding. The encoding is implemented by switching between different components of $\Theta_j$, while the decoding uses a reservoir computer and the training coefficients ${\bf W}^{out}_j$ from eq. (\ref{trainfit}) to determine which component was transmitted. If there are $N_s$ possible components of $\Theta$ that can be transmitted, then the number of bits of information in each data interval is ${\rm log}_2(N_s)$. The detection may be done coherently (using a synchronized receiver) or non-coherently (using an asynchronous receiver).

The data signal consists of a series of discreet values ${\mathcal I}(k), k=1 \ldots N_d$, where ${\mathcal I}(k)$ is an integer in the range 1 to $N_s$. In the $k$'th time slot, the signal $\Theta_{{\mathcal I}(k)}$ is transmitted, while in the $k+1$ time slot the signal to be transmitted is $\Theta_{{\mathcal I}(k+1)}$. The transmission between the $\Theta$ signals for the two time intervals may not be continuous, so there is a transition interval during which both the $k$'th and $k+1$'th signals are multipled by a variable weight.

The weighting factor used to smooth the transition between communications symbols is
\begin{equation}
\label{wfact}
{\mathcal W}\left(n\right)=
\left\{ \begin{array}{cc}
n/N_{b} & n\leq N_{b}\\
1 & N_{b}<n\leq {\mathcal N}_{I}
\end{array}\right\} 
\end{equation}

where $N_b$ is the breakpoint and the time slot contains ${\mathcal N}_I$ points. For all the simulations reported here, $N_b = {\mathcal N}_I/4$. For time slot $k$ the encoded signal $S^I(k)$ is
\begin{equation}
\label{encsig}
S^I(k)={\mathcal W} \times \Theta_{{\mathcal I}(k)}\left ( n \ldots n+{\mathcal N}_I \right ) + (1-{\mathcal W}) \times \Theta_{{\mathcal I}(k-1)} \left ( n \ldots n+{\mathcal N}_I \right ) .
\end{equation}
In the $k+1$'th time slot the encoded signal is
\begin{equation}
\label{encsig1}
S_{I}(k+1)={\mathcal{W}}\times\Theta_{{\mathcal{I}}(k+1)}\left(n+{\mathcal{N}}_{I}\ldots n+2{\mathcal{N}}_{I}\right)+(1-{\mathcal{W}})\times\Theta_{{\mathcal{I}}(k)}\left(n+{\mathcal{N}}_{I}\ldots n+2{\mathcal{N}}_{I}\right).
\end{equation}

The transmitted signal $S^T$ is
\begin{equation}
\label{transsig}
S^T=S^I+\eta
\end{equation}
where $\eta$ is an additive Gaussian white noise signal.

\subsection{Synchronous Decoding}
For synchronous decoding the signal ${\tilde s}$ that drives the original reservoir in eq. \ref{tanhnode} is available at the receiver. The chaotic systems that produced ${\tilde s}$ may be synchronized using a synchronization preamble prepended to the transmitted signal , or the initial conditions may be stored as a key. In \cite{carroll2017b} a chaotic receiver was synchronized using a stored library of chaotic sequences and a correlation receiver.

For synchronous decoding a reservoir computer identical to eq. (\ref{tanhnode}) is driven with the synchronized version of ${\tilde s}$ and the outputs are arranged in a matrix $\tilde{\Omega}$ analogous to $\Omega_1$ in eq. (\ref{newmat}). A set of $N_s$ chaotic signals is then produced as $\tilde{\Theta}=\tilde{\Omega} \Psi_O$, where $\Psi_O$ is the same set of coefficients used in eq. (\ref{randsig}). The testing target signals are $\tilde {g}_j(n)=\tilde{\Theta}_j(n+1)$, the signals from $\tilde {\Theta}_j$ predicted one time step into the future.

To compare the testing targets to the transmitted signal, a reservoir computer identical to the one in eq. (\ref{resdid}) is used to identify which signal from $\Theta_j, j=1 \ldots N_s$ was actually transmitted. The reservoir is driven by the transmitted signal $S^T$ and the output signals are arranged in a matrix $\Omega^T$ as in eq. (\ref{fitmat}). To compare to each possible chaotic signal from the set $\Theta$, the output matrix in the receiver, $\Omega^T$ , is multiplied by the trained coefficients ${\bf W}^{out}_j$ that were found in eq. (\ref{train_fit}) for each signal in $\Theta$. The testing errors are then calculated as
\begin{equation}
\label{testerr}
\Delta_{r}\left(j\right)=\sum_{n=1}^{N_{I}}\tilde{\Theta}_{j}\left(n+1\right)-\Omega_{l}^{T}\left(n\right)\mathbf{W}_{j}^{out}
\end{equation}
where the sum is taken over the $k$'th data interval.

 The detected symbol is
\begin{equation}
\varsigma_{det} = \underset{j}{\arg\min} \: \Delta_r(j).
\end{equation}
If $\varsigma_{det}$ is not the actual symbol that was transmitted, then an error is recorded. The fraction of symbols that are incorrect is $P_e$.

\subsection{Asynchronous (non-coherent) Decoding}
It may be that the signal ${\tilde s}$ that drove the original reservoir, ${\tilde s}$, is not available at the receiver, in which case decoding must take place asynchronously. For asynchronous decoding, the testing target is the value of the transmitted signal one time step into the future: $\tilde {g}(n)=S^T(n+1)$.  Unlike synchronous decoding, the testing target is contaminated with noise. As with synchronous decoding, a reservoir identical to eq. (\ref{resdid}) is driven by the signal $S^T$ and the output signals are arranged in a matrix $\Omega^T$ as in eq. (\ref{fitmat}). As in the synchronous case, the matrix $\Omega^T$ , is multiplied by the trained coefficients ${\bf W}^{out}_j$ that were found in eq. (\ref{trainfit}). In the asynchronous case, the detection errors are
\begin{equation}
\label{asyncerr}
\Delta_{r}\left(j\right)=\sum_{n=1}^{N_{I}}S^{T}\left(n+1\right)-\Omega_{l}^{T}\left(n\right)\mathbf{W}_{j}^{out}.
\end{equation}

\section{Entropy Statistic}
 Obtaining the maximum diversity of signals in the reservoir computer should make it easier to distinguish the different chaotic signals produced in eq. (\ref{randsig}).  Entropy is a useful way to characterize this diversity. It was shown in \cite{carroll2021b} that classification of different signals was optimal when the entropy of the reservoir was maximized. Measuring entropy requires a partitioning of the dynamical system. In \cite{xiong2017} there are a number of ways to do this partitioning, although different partitions can give different results for the entropy.  It was found that the permutation entropy method \cite{bandt2002} avoided this coarse graining problem because it creates partitions based on the time ordering of the signals.  Each individual node time series $r_i(n)$  was divided into windows of 4 points, and the points within the window were sorted to establish their order; for example, if the points within a window were 0.1, 0.3, -0.1 0.2, the ordering would be 3, 1, 4, 2. Each possible ordering of points in a signal $r_i(n)$ represented a symbol $\psi_i(n)$.
  
At each time step $n$, the individual node signals were combined into a reservoir computer symbol $\Lambda(n)=[\psi_1(n), \psi_2(n), \ldots \psi_M(n)]$. For a large number of nodes there were potentially a huge number of possible symbols, but the nodes were all driven by a common drive signal, so only a tiny fraction of the symbol space was actually occupied, on the order of tens of symbols for the entire reservoir computer.

If $Q$ total symbols were observed for the reservoir computer for the entire time series, then the reservoir computer entropy was
\begin{equation}
\label{entropy}
H =  - \sum\limits_{q = 1}^Q {p\left( {{\Lambda _q}} \right)\log \left( {p\left( {{\Lambda _q}} \right)} \right)} 
\end{equation}
where $q(\Lambda_q)$ is the probability of the $q$'th symbol.

\section{Results}
The driving signal in the transmitter is the $x$ signal from the chaotic Lorenz system. The Lorenz system is described by \cite{lorenz1963}
\begin{equation}
\label{loreq}
\begin{array}{l}
\frac{{dx}}{{dt}} =T_l \left( {p_1}\left( {y - x} \right) \right)\\
\frac{{dy}}{{dt}} = T_l \left( x\left( {{p_2} - z} \right) - y\right)\\
\frac{{dy}}{{dt}} = T_l \left( xy - {p_3}z\right)
\end{array}
\end{equation}
with $p_1=10$, $p_2=28$, $p_3=8/3$ and $T_l=0.1$. The Lorenz equations were numerically integrated with a time step of 1. The input to the reservoir computer was $x$, and the reservoir computer was trained to output $z$. The $x$ signal was normalized to have a mean of zero and a standard deviation of 1.

\begin{figure}
\centering
\includegraphics[scale=0.8]{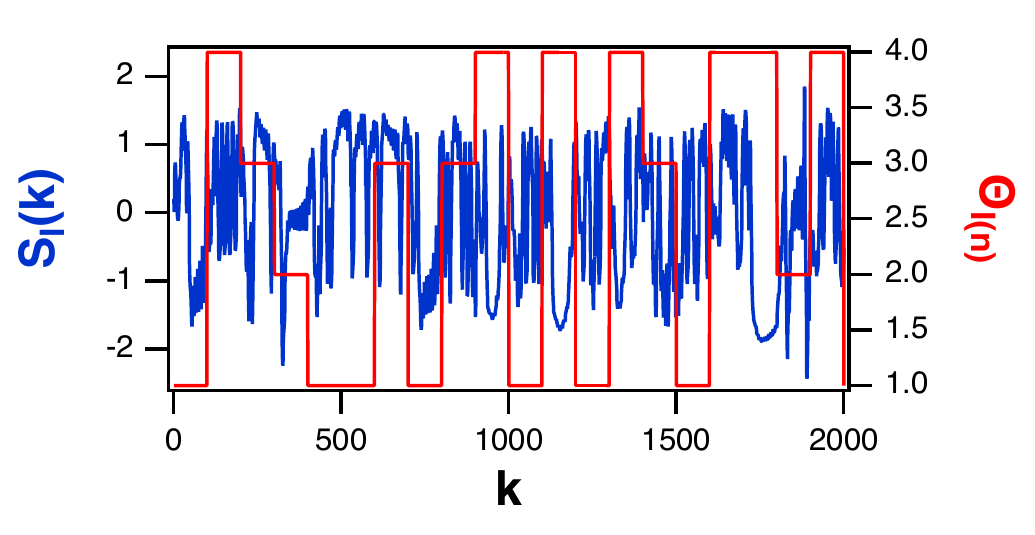}
\caption{\label{datasignals} In blue is the encoded signal that was switched between one of four different signals every 100 time steps. In red is the encoded data value. The weighting factor of eq. (\ref{wfact}) smoothed the transition between signals.}
\end{figure}

The reservoir computer of eq. (\ref{tanhnode}) with $M$ nodes was driven by the Lorenz $x$ signal to produce $N_s$ new signals as in eq. (\ref{randsig}). As a first step the parameter $\alpha$ and the spectral radius $\sigma$ of the adjacency matrix were swept to determine the best values for these parameters. The spectral radius is the largest absolute value of the eigenvalues of the adjacency matrix ${\bf A}$. For these simulations the reservoir computer had $M=50$ nodes and was used to produce $N_s=4$ signals. The length of one data interval was ${\mathcal N}_I=100$ points, and the transmitted signal had a standard deviation of 1.  Figure \ref{datasignals} shows both the encoded $S^I$ and the encoded data values  $\Theta_{{\mathcal I}(n)}$. 

\begin{figure}
\centering
\includegraphics[scale=0.8]{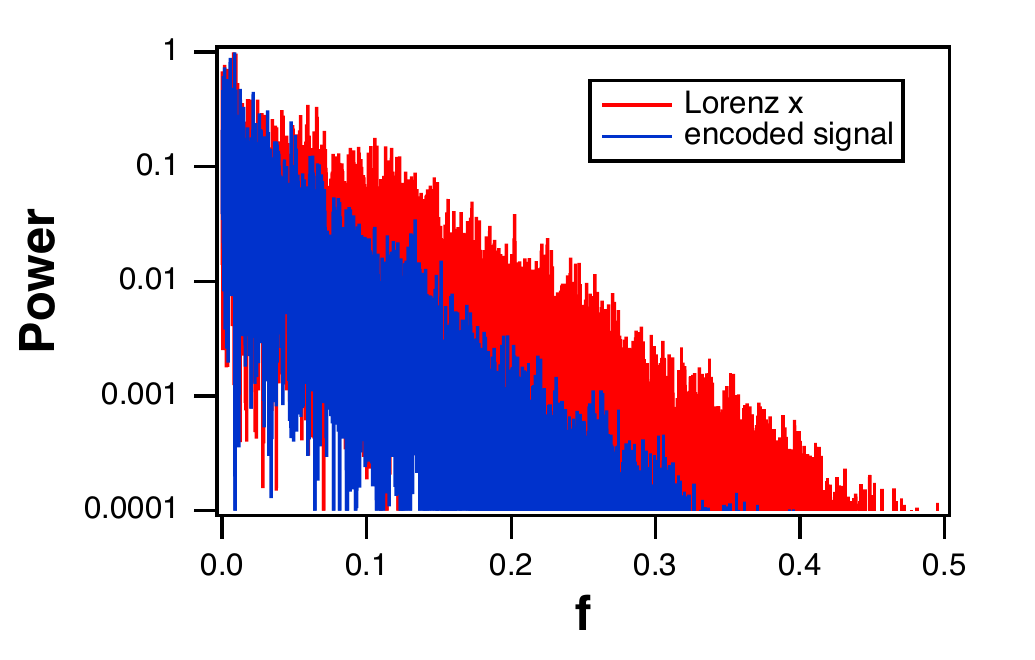}
\caption{\label{signalspect} Power spectra of the Lorenz $x$ signal and the encoded signal $S^I$.  }
\end{figure}

Figure \ref{signalspect} shows the power spectra of the Lorenz $x$ signal and the encoded signal $S^I$. The encoded spectrum drops off faster than the Lorenz spectrum, but because of the weighting factor used to smooth the transition between different communications symbols, the spectrum does not contain a signature of this switching.

\begin{figure}
\centering
\includegraphics[scale=0.8]{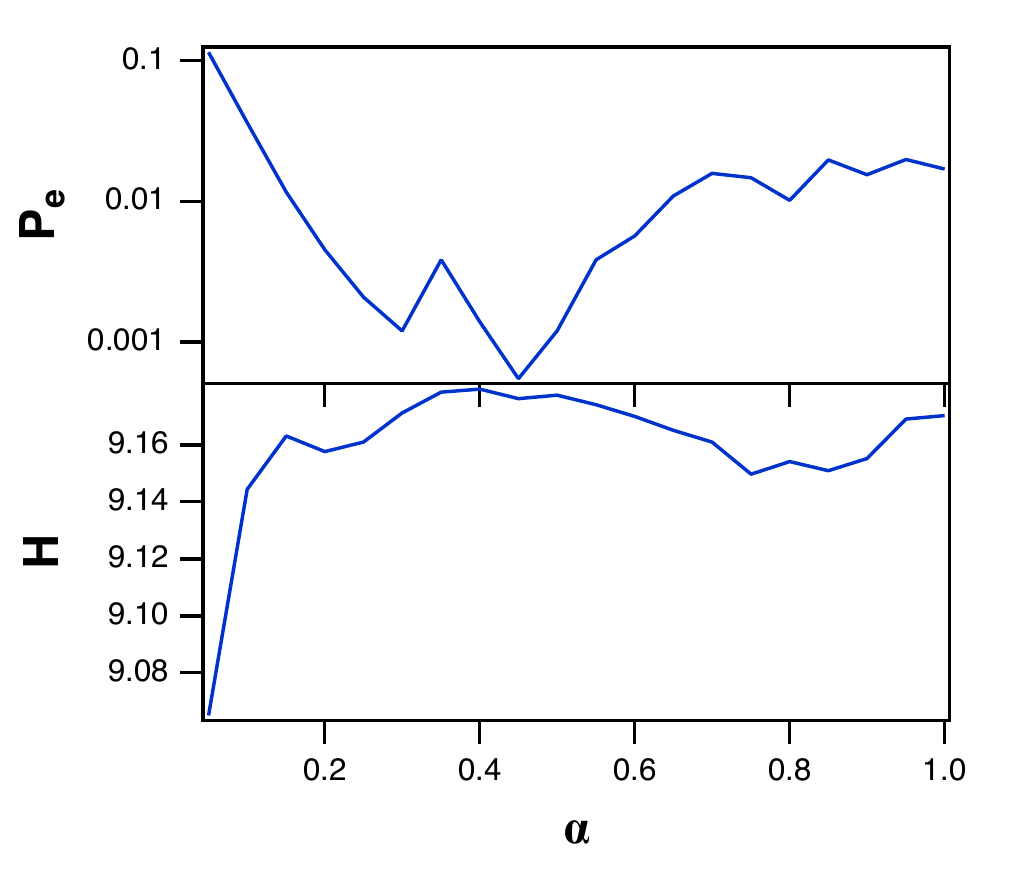}
\caption{\label{alphavar} The top plot is the probability of making an error $P_e$ in determining which one of four different chaotic signals was transmitted as the 50 node reservoir computer parameter $\alpha$ was varied while the spectral radius $\sigma=1$. The receiver was operated in the synchronous configuration, and added noise standard deviation was half the signal standard deviation. Each data interval contained 100 time steps. The bottom plot is the entropy H of the reservoir as calculated according to eq. (\ref{entropy})}.
\end{figure}

 The detection was done in the synchronous configuration, but for these parameters there were no errors for the detected signal, so Gaussian random noise with a standard deviation of 0.5 was added to the transmitted signal. The communications system was simulated 500 times for each parameter value, and each trial simulated a transmission with 1000 data intervals of length ${\mathcal N}_I=100$ points each. The probability of error $P_e$ as a function of $\alpha$ with these parameters is shown in figure \ref{alphavar}.

Figure \ref{alphavar} shows that the minimum probability of error coincides with the maximum in the entropy of the reservoir. This result is expected, as it was shown in \cite{carroll2021b} that the optimum classification performance for reservoir computers came when their entropy was maximized.

\begin{figure}
\centering
\includegraphics[scale=0.8]{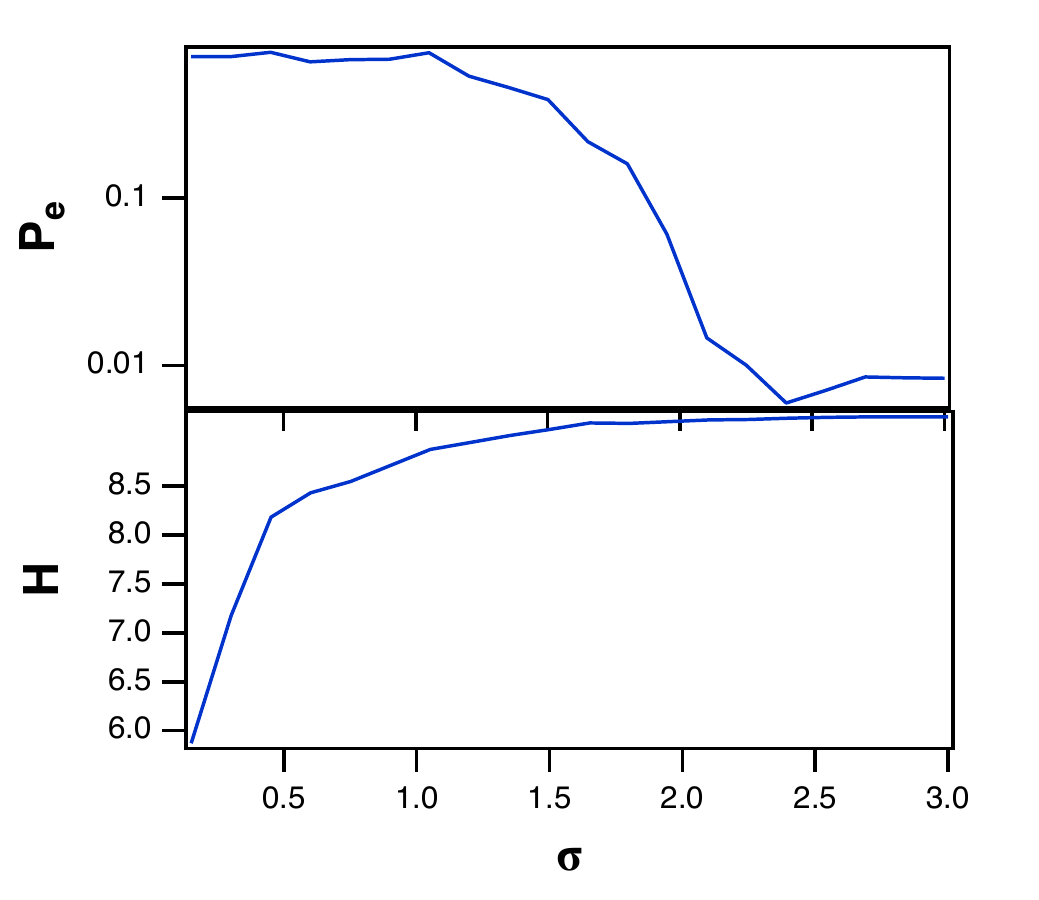}
\caption{\label{spradvar} The top plot is the probability of making an error $P_e$ in determining which one of four different chaotic signals was transmitted as the spectral radius $\sigma$ for the 50 node reservoir computer was varied while the $\alpha=0.45$. The receiver was operated in the synchronous configuration, and added noise standard deviation was half the signal standard deviation. Each data interval contained 100 time steps. The bottom plot is the entropy H of the reservoir as calculated according to eq. (\ref{entropy})}.
\end{figure}

The probability of error as the adjacency matrix spectral radius $\sigma$ is varied is plotted in figure \ref{spradvar}.

\begin{figure}
\centering
\includegraphics[scale=0.8]{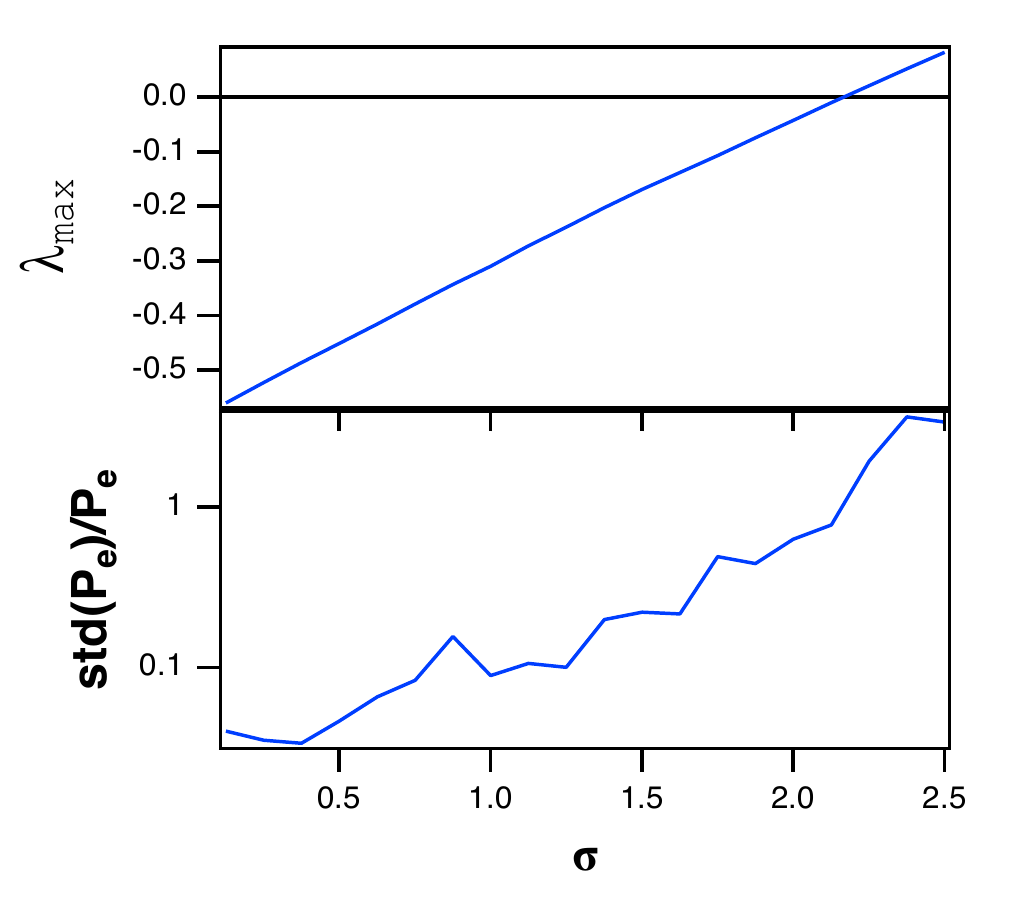}
\caption{\label{spradlya} The top plot is the maximum Lyapunov exponent for the reservoir as the spectral radius $\sigma$ for the 50 node reservoir computer was varied while the parameter $\alpha=0.45$. For the largest values of $\sigma$, the reservoir was actually chaotic. The bottom plot is the standard deviation in the probability of error normalized by the probability of error $P_e$.  }
\end{figure}

Figure \ref{spradvar} also shows that the best performance for distinguishing the different chaotic signals comes when the reservoir computer entropy is large, but there is a complication in using large values of the spectral radius. Figure \ref{spradlya} shows the maximum Lyapunov exponent for the reservoir computer and the standard deviation of the probability of error normalized by the probability of error $P_e$. The Lyapunov exponent was calculated by the Gram-Schmidt method.

Figure \ref{spradlya} shows that for the largest values of the spectral radius the reservoir became chaotic. This would seem to contradict figure \ref{spradvar}, where the lowest probability of error also came at the largest values of the spectral radius. If the goal was to fit or predict a signal, then chaos would be detrimental; the goal here, however is to classify signals. As long as the response of the reservoir computer is distinctly different for each of the different possible input signals, chaos does not increase the probability of classification error.

The chaotic behavior does cause some problems, however. The bottom plot in figure \ref{spradlya} shows the ratio of the standard deviation of the probability of error to the mean probability of error. The ratio increases as the spectral radius increases, so that for the largest values of the spectral radius, the standard deviation in the error probability is las large as the mean probability itself. For the largest spectral radii, the probability of error in the receiver was not be reproducible; sometimes it would be large and sometimes it would be small. As a result, in the following simulations, the spectral radius will be set to $\sigma=1$. 

\begin{figure}
\centering
\includegraphics[scale=0.8]{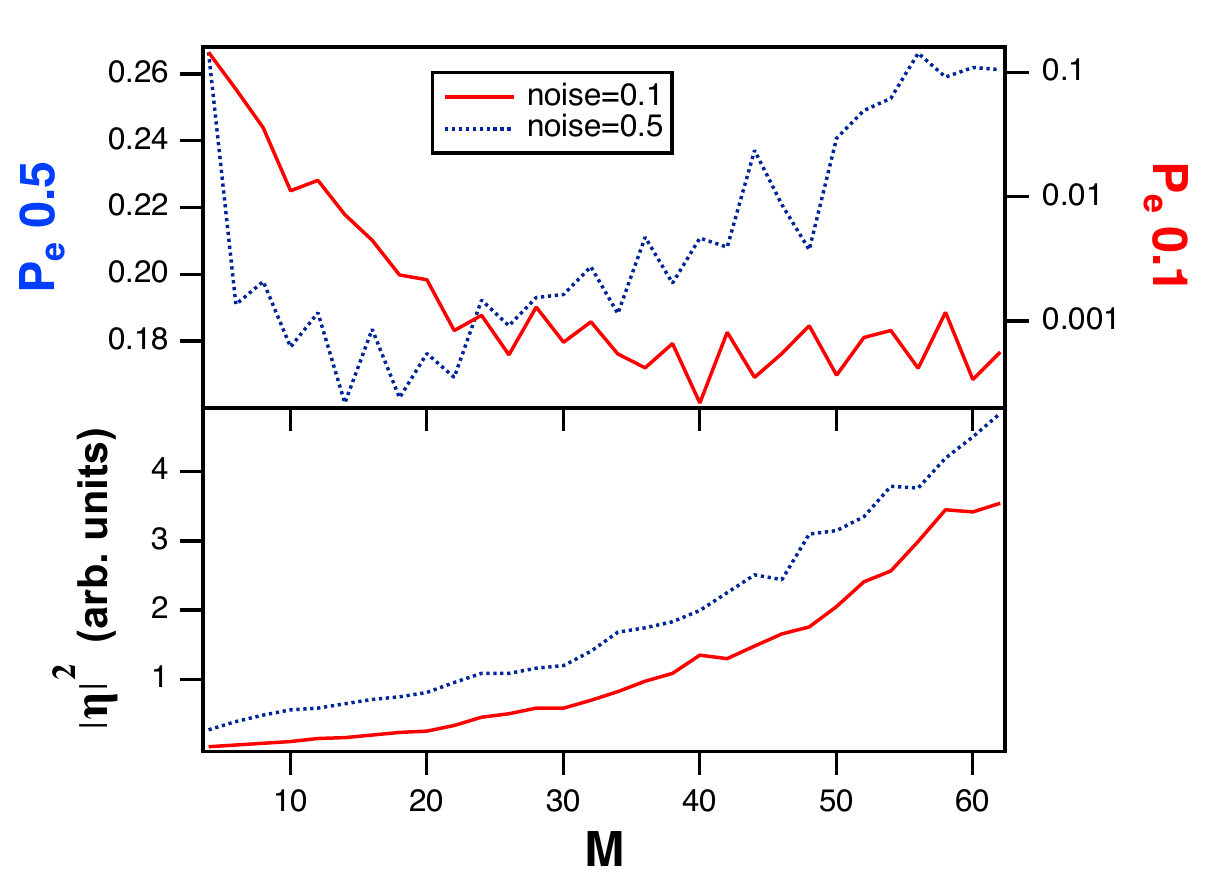}
\caption{\label{nodevar} The top plot is the probability of error $P_e$ in detecting the different chaotic signals as the number of nodes $M$ in the reservoir computer changed. The receiver was operated in the synchronous configuration. For the solid red line (right axis), the added noise had an amplitude of 0.1, while the dotted blue line (left axis) represents an added noise amplitude of 0.1. The bottom plot shows the total noise power in the fitted signals from a reservoir driven by noise only, $\left | \eta \right | ^2$, for both noise amplitudes. Contrary to expectations, for the higher noise amplitude the probability of error does not decrease as the number of nodes increases.}
\end{figure}

In most reservoir computer simulations, constructing larger reservoir computers improves performance, but constructing larger reservoir computers also increases the system complexity and cost, so it is useful to know if larger reservoir computers give a smaller probability of error. The number of nodes in the reservoir computer were varied from 4 to 61 while other parameters were held constant. For this simulation there were again four signals and the length of each data interval was 100 time steps. The noise standard deviation was 0.5. The parameter $\alpha$ was fixed at 0.45 while the spectral radius $\sigma$ for the adjacency matrix was 1. The top plot in figure \ref{nodevar} shows the variation in the probability of error as the number of nodes changed.

The top plot in figure \ref{nodevar} shows that for a low noise level the probability of detection error $P_e$ increases as the level of added noise increases, but for a higher noise level the error probability goes through a minimum and actually starts to increase as the number of nodes increases.  The bottom plot in fig. \ref{nodevar} shows why the probability of error can increase as the number of nodes increases.

The bottom plot in fig. \ref{nodevar} was created by driving a reservoir computer with Gaussian white noise only. The reservoir output signals $r_{ij}(n), i=1 \ldots M, j=1 \ldots N_s$ for each of $N_s=4$ different signals were multiplied by the corresponding training coefficients ${\bf W}^{out}_j$ to create four fit signals $h_j=\Omega_j {\bf W}^{out}_j$.The total power in each of the signals $h_j$ was summed and normalized by the number of points in the time series to create the plot in the bottom of fig. \ref{nodevar}.

 When noise is added to the input signal $S^T$ (eq. \ref{transsig}), the noise as well as the signal is multiplied by the input vector ${\bf W}^{in}$ before being added to each node. The same noise is added to each node, so the noise signals on different nodes will be correlated. When the linear fit to a training or testing signal is made as in eq. (\ref{trainfit}), the correlated noise signals from each node are added together, so the noise amplitude is multiplied. The nonlinear nature of the reservoir makes it difficult to specify this multiplication factor, but the bottom plot in fig. \ref{nodevar} shows a monotonic increase in the noise power as the number of nodes is increased. When the noise level is small (noise = 0.1), increasing the number of nodes overcomes the increase in output noise, so the probability of error can decrease, but for a larger noise amplitude (noise = 0.5) the increased number of nodes does not compensate for the larger output noise. Fig. \ref{nodevar} illustrates a compromise in using a reservoir computer to process noisy signals; increasing the size of the reservoir can also increase the errors caused by added noise.

\subsection{Performance versus Noise}
The  specific set of random coefficients $\Psi_O$ could affect the ability to distinguish the different signals, so a Monte Carlo simulation was used to find a set of coefficients that produced the lowest probability of errors in classifying the signals. For every simulation with a fixed number of signals $N_s$ and a number of nodes $M$, 400 different sets of coefficients were generated randomly and the error fraction $P_e$ was determined for each set of coefficients. The coefficients the smallest $P_e$ were then used in further communications simulations in which the noise level or length of the data interval were changed .

\begin{figure}
\centering
\includegraphics[scale=0.8]{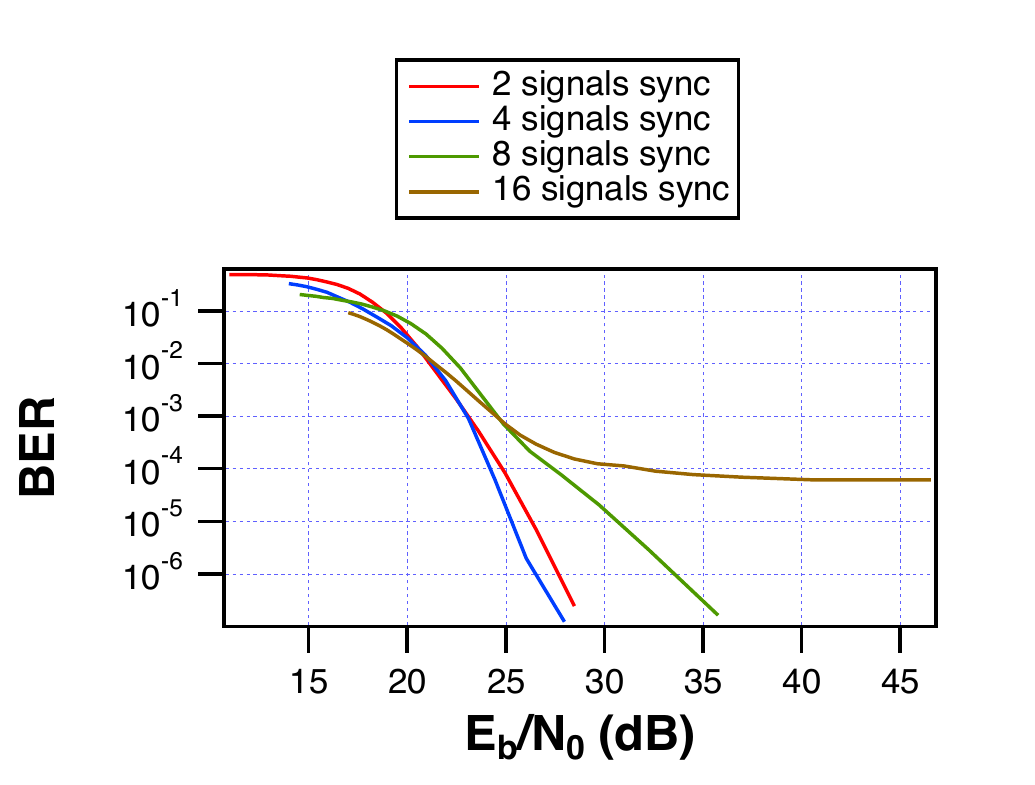}
\caption{\label{noisebersync} Bit error rate (BER) for the synchronous receiver as a function of energy per bit normalized by the power spectral density ($E_b/N_0$).}
\end{figure}

\begin{figure}
\centering
\includegraphics[scale=0.8]{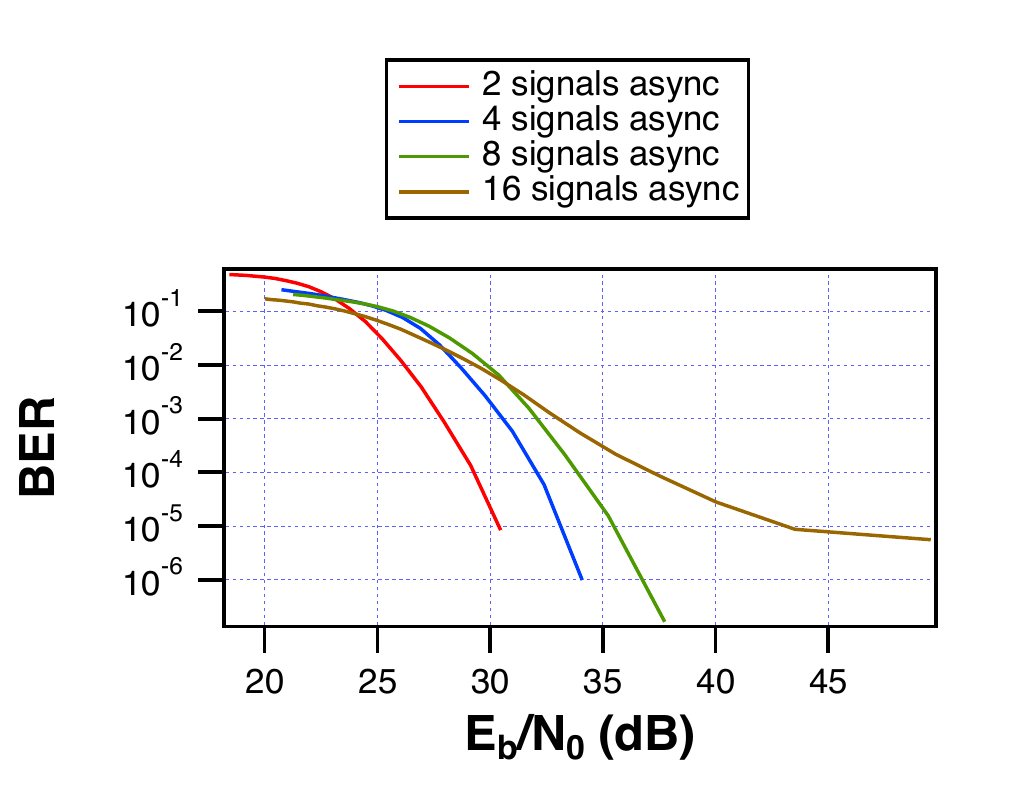}
\caption{\label{noiseberasync} Bit error rate (BER) for the asynchronous receiver as a function of energy per bit normalized by the power spectral density ($E_b/N_0$).}
\end{figure}

Communications system performance is usually characterized in terms of the bit error rate (BER) as a function of the energy per bit normalized by the noise power spectral density ($E_b/N_0$). The probability of error $P_e$ calculated in the previous sections is the probability of making an error in determining which symbol is present. If there are $N_s$ different symbols, then ${\rm log}_2(N_s)$ bits are sent during each data interval. The BER may be approximated by dividing the probability of error by the number of bits in the interval: $\mathrm{BER}\approx P_{e}\mathrm{/log_{2}}\left(N_{s}\right) $ The BER as a function of $E_b/N_0$ for both synchronous and asynchronous configurations was measured as the level of additive Gaussian white noise was increased. For the synchronous configuration, the length of one data inverval was ${\mathcal N}_I=100$, $\alpha=0.45$, the spectral radius was $\sigma=1$ and the reservoir had $M=50$ nodes. For the asynchronous configuration the parameters were the same except that the length of one data interval was ${\mathcal N}_I=400$. The results for the synchronous receiver are in  figure \ref{noisebersync} and for the asynchronous receiver in \ref{noiseberasync}.

For large noise levels (smaller $E_b/N_0$) the bit error rates for different numbers of signals are about the same. The receiver is nonlinear, meaning that added noise in the communications channel gets mixed in with the communications symbols, making them impossible to distinguish at high noise levels. For lower noise levels, the bit error rate is higher for larger numbers of signals. In standard linear receivers for antipodal communications, such as binary or quadrature phase shift keying, signals carrying different numbers of bits lie on the same BER vs. $E_b/N_0$ curve, but the receiver here is nonlinear and the different communications symbols are not orthogonal, so transmitting more bits in the same interval increases the bit error rate.

As the number of signals becomes larger in figure \ref{noisebersync} the bit error rate curves approach an asymptote instead of continuing to tail off. For the synchronous receiver with 16 signals and a data interval of ${\mathcal N}_I=100$, even with zero noise the bit error rate is approximately $10^{-4}$. For the asynchronous receiver with the same parameters and  no noise the probability of error is approximately $4 \times 10^{-3}$. 

The BER vs. $E_b/N_0$ curves for the asynchronous receiver in figure \ref{noiseberasync} show a larger difference as the number of signals increases. For 2 signals, to achieve a BER of 0.01, the asynchronous receiver needs an $E_b/N_0$ 5 dB larger than the synchronous receiver.
 
\section{Conclusions}
One difficulty in applications of chaotic signals is the difficulty in designing continuous chaotic systems. This work demonstrates how to use a reservoir computer to generate multiple chaotic signals with similar spectra from a known chaotic system. A simple communications application was used to demonstrate that these different chaotic signals were easily distinguishable. It was also shown that attempting to improve the detection performance for these different chaotic signals by increasing the size of the reservoir computer could actually make performance worse by effectively multiplying noise.

The synchronous version of this application may also be suitable for radar. The reservoir computer is in a state of generalized synchronization with the driving Lorenz signal, so the $N_s$ new chaotic signals that are created are correlated with each other. As a result, the correlation between the Lorenz driving signal and the transmitted signal will be little affected by the presence of a message in the transmitted signal.

In this work the reservoir used to create the new chaotic signals and the reservoir used to distinguish them were identical, but they could be different. Using the same reservoir for creating the new chaotic signals and classifying them is convenient because only one reservoir is needed, while different coefficients are used for different tasks, but using multiple reservoirs is possible. One could use a larger reservoir to create the new chaotic signals and a smaller reservoir to distinguish them to avoid some of the problems with driving large reservoirs with noisy signals. The node types or parameters could also be different.

The different chaotic signals generated in these simulations are correlated with each other, but a set of orthogonal signals could be produced from these signals, as was done in \cite{carroll2017b} or \cite{carroll2017c}, although the signal spectra may be altered by the process of creating orthogonal signals.

\section{Acknowledgements}
This work was supported by the Naval Research Laboratory's Basic Research Program.

\section*{References}

\bibliography{reservoircsk}

\end{document}